\documentclass[final,10pt,twocolumn]{IEEEtran}
\usepackage{amsmath,amssymb,graphicx,psfrag,cite,url}

\usepackage[caption=false,font=footnotesize]{subfig}

\newcommand{\bc}{\begin{center}}
\newcommand{\ec}{\end{center}}

\newcommand{\ba}{\begin{array}}
\newcommand{\ea}{\end{array}}

\newcommand{\bt}{\begin{tabular}}
\newcommand{\et}{\end{tabular}}

\newcommand{\eems}{\end{multline*}}
\newcommand{\eem}{\end{multline}}
\newcommand{\eeams}{\end{eqnarray}\vspace{0.1cm}}
\newcommand{\eearms}{\end{eqnarray*}\vspace{0.1cm}}
\newcommand{\edmms}{\end{displaymath}\vspace{0.1cm}}

\newcommand{\bems}{\begin{multline*}}
\newcommand{\bem}{\begin{multline}}
\newcommand{\beams}{\vspace{0.1cm} \begin{eqnarray}}
\newcommand{\bearms}{\vspace{0.1cm} \begin{eqnarray*}}
\newcommand{\bdmms}{\vspace{0.1cm} \begin{displaymath}}

\newcommand{\be}{\begin{equation}}
\newcommand{\ee}{\end{equation}}

\newcommand{\bes}{\begin{equation*}}
\newcommand{\ees}{\end{equation*}}

\newcommand{\bea}{\begin{eqnarray}}
\newcommand{\eea}{\end{eqnarray}}

\newcommand{\bear}{\begin{eqnarray*}}
\newcommand{\eear}{\end{eqnarray*}}

\newcommand{\bdm}{\begin{displaymath}}
\newcommand{\edm}{\end{displaymath}}

\newcommand{\feop}{\hfill \rule{1.5mm}{1.5mm} \\}

\newcommand{\bit}{\begin{itemize}}
\newcommand{\eit}{\end{itemize}}

\newcommand{\ben}{\begin{enumerate}}
\newcommand{\een}{\end{enumerate}}

\newcommand{\bd}{\begin{document}}
\newcommand{\ed}{\end{document}}

\newcommand{\psp}{\hspace{0.6cm}}
\newcommand{\hspone}{\hspace{1cm}}
\newcommand{\hsptwo}{\hspace{2cm}}

\newcommand{\hspthree}{\hspace{3cm}}
\newcommand{\hspfour}{\hspace{4cm}}

\newcommand{\hspfive}{\hspace{5cm}}
\newcommand{\hspsix}{\hspace{6cm}}

\newcommand{\hspseven}{\hspace{7cm}}
\newcommand{\hspeight}{\hspace{8cm}}

\newcommand{\vspone}{\vspace{1cm}}
\newcommand{\vsptwo}{\vspace{2cm}}

\newcommand{\vspthree}{\vspace{3cm}}
\newcommand{\vspfour}{\vspace{4cm}}

\newcommand{\vspfive}{\vspace{5cm}}
\newcommand{\vspsix}{\vspace{6cm}}

\newcommand{\vspseven}{\vspace{7cm}}
\newcommand{\vspeight}{\vspace{8cm}}

\newcommand{\vspnine}{\vspace{9cm}}
\newcommand{\vspten}{\vspace{10cm}}

\newcommand{\vspeleven}{\vspace{11cm}}
\newcommand{\vsptwelve}{\vspace{12cm}}

\newcommand{\qett}{\mbox{$(q^{-1})$}}
\newcommand{\zett}{\mbox{$(z^{-1})$}}

\newcommand{\qo}{q^{-1}}
\newcommand{\zettu}{\mbox{$z^{-1}$}}

\newcommand{\linf}[1]{{#1}\rightarrow \infty}

\newcommand{\cf}[1]{(\ref{#1})}

\newcommand{\eiw} {\mathrm{e}^{\mathrm{i}\omega}}
\newcommand{\emiw}{\mathrm{e}^{-\mathrm{i}\omega}}

\newcommand{\tr}{\mbox{${\rm Tr}$}}
\newcommand{\var}{\mbox{${\rm Var}$}}
\newcommand{\diag}{\mbox{${\rm Diag}$}}
\newcommand{\cov}{\mbox{${\rm Cov}$}}
\newcommand{\Ex}{{\sf E}}

\newcommand{\erm}{\mathrm{e}}
\newcommand{\irm}{\mathrm{i}}
\newcommand{\Prj} {{\bf \Pi}^\bot}

\newcommand{\bA} {{\bf A}}
\newcommand{\bB} {{\bf B}}
\newcommand{\bC} {{\bf C}}
\newcommand{\bD} {{\bf D}}
\newcommand{\bE} {{\bf E}}
\newcommand{\bF} {{\bf F}}
\newcommand{\bG} {{\bf G}}
\newcommand{\bH} {{\bf H}}
\newcommand{\bI} {{\bf I}}
\newcommand{\bJ} {{\bf J}}
\newcommand{\bK} {{\bf K}}
\newcommand{\bL} {{\bf L}}
\newcommand{\bM} {{\bf M}}
\newcommand{\bN} {{\bf N}}
\newcommand{\bO} {{\bf O}}
\newcommand{\bP} {{\bf P}}
\newcommand{\bQ} {{\bf Q}}
\newcommand{\bR} {{\bf R}}
\newcommand{\bS} {{\bf S}}
\newcommand{\bT} {{\bf T}}
\newcommand{\bU} {{\bf U}}
\newcommand{\bV} {{\bf V}}
\newcommand{\bW} {{\bf W}}
\newcommand{\bY} {{\bf Y}}
\newcommand{\bX} {{\bf X}}
\newcommand{\bZ} {{\bf Z}}

\newcommand{\GLm} {{\bf \Lambda}}
\newcommand{\GSg} {{\bf \Sigma}}
\newcommand{\GPh} {{\bf \Phi}}
\newcommand{\GPs} {{\bf \Psi}}
\newcommand{\GPi} {{\bf \Pi}}
\newcommand{\GDl} {{\bf \Delta}}
\newcommand{\GGm} {{\bf \Gamma}}
\newcommand{\GUp} {{\bf \Upsilon}}
\newcommand{\GOm} {{\bf \Omega}}
\newcommand{\GTh} {{\bf \Theta}}
\newcommand{\GXi} {{\bf \Xi}}

\newcommand{\gal} {\boldsymbol{\alpha}}
\newcommand{\gbt} {\boldsymbol{\beta}}
\newcommand{\ggm} {\boldsymbol{\gamma}}
\newcommand{\gdl} {\boldsymbol{\delta}}
\newcommand{\gep} {\boldsymbol{\epsilon}}
\newcommand{\vep} {\boldsymbol{\varepsilon}}
\newcommand{\gzt} {\boldsymbol{\zeta}}
\newcommand{\get} {\boldsymbol{\eta}}
\newcommand{\gth} {\boldsymbol{\theta}}
\newcommand{\vth} {\boldsymbol{\vartheta}}
\newcommand{\git} {\boldsymbol{\iota}}
\newcommand{\gkp} {\boldsymbol{\kappa}}
\newcommand{\glm} {\boldsymbol{\lambda}}
\newcommand{\gmu} {\boldsymbol{\mu}}
\newcommand{\gnu} {\boldsymbol{\nu}}
\newcommand{\gxi} {\boldsymbol{\xi}}
\newcommand{\gpi} {\boldsymbol{\pi}}
\newcommand{\vpi} {\boldsymbol{\varpi}}
\newcommand{\grh} {\boldsymbol{\rho}}
\newcommand{\vrh} {\boldsymbol{\varrho}}
\newcommand{\gsg} {\boldsymbol{\sigma}}
\newcommand{\vsg} {\boldsymbol{\varsigma}}
\newcommand{\gtu} {\boldsymbol{\tau}}
\newcommand{\gup} {\boldsymbol{\upsilon}}
\newcommand{\gph} {\boldsymbol{\phi}}
\newcommand{\vph} {\boldsymbol{\varphi}}
\newcommand{\gch} {\boldsymbol{\chi}}
\newcommand{\gps} {\boldsymbol{\psi}}
\newcommand{\gom} {\boldsymbol{\omega}}

\newcommand{\Ba}  {{\bf a}}
\newcommand{\Bb}  {{\bf b}}
\newcommand{\Bc}  {{\bf c}}
\newcommand{\Bd}  {{\bf d}}
\newcommand{\Be}  {{\bf e}}
\newcommand{\Bf}  {{\bf f}}
\newcommand{\Bg}  {{\bf g}}
\newcommand{\Bh}  {{\bf h}}
\newcommand{\Bi}  {{\bf i}}
\newcommand{\Bj}  {{\bf j}}
\newcommand{\Bk}  {{\bf k}}
\newcommand{\Bl}  {{\bf l}}
\newcommand{\Bm}  {{\bf m}}
\newcommand{\Bn}  {{\bf n}}
\newcommand{\Bo}  {{\bf o}}
\newcommand{\Bp}  {{\bf p}}
\newcommand{\Bq}  {{\bf q}}
\newcommand{\Br}  {{\bf r}}
\newcommand{\Bs}  {{\bf s}}
\newcommand{\Bt}  {{\bf t}}
\newcommand{\Bu}  {{\bf u}}
\newcommand{\Bv}  {{\bf v}}
\newcommand{\Bw}  {{\bf w}}
\newcommand{\Bx}  {{\bf x}}
\newcommand{\By}  {{\bf y}}
\newcommand{\Bz}  {{\bf z}}

\newcommand{\za} {\boldsymbol{a}}
\newcommand{\zb} {\boldsymbol{b}}
\newcommand{\zc} {\boldsymbol{c}}
\newcommand{\zd} {\boldsymbol{d}}
\newcommand{\ze} {\boldsymbol{e}}
\newcommand{\zf} {\boldsymbol{f}}
\newcommand{\zg} {\boldsymbol{g}}
\newcommand{\zh} {\boldsymbol{h}}
\newcommand{\zi} {\boldsymbol{i}}
\newcommand{\zj} {\boldsymbol{j}}
\newcommand{\zk} {\boldsymbol{k}}
\newcommand{\zl} {\boldsymbol{\ell}}
\newcommand{\zm} {\boldsymbol{m}}
\newcommand{\zn} {\boldsymbol{n}}
\newcommand{\zo} {\boldsymbol{o}}
\newcommand{\zp} {\boldsymbol{p}}
\newcommand{\zq} {\boldsymbol{q}}
\newcommand{\zr} {\boldsymbol{r}}
\newcommand{\zs} {\boldsymbol{s}}
\newcommand{\zt} {\boldsymbol{t}}
\newcommand{\zu} {\boldsymbol{u}}
\newcommand{\zv} {\boldsymbol{v}}
\newcommand{\zw} {\boldsymbol{w}}
\newcommand{\zx} {\boldsymbol{x}}
\newcommand{\zy} {\boldsymbol{y}}
\newcommand{\zz} {\boldsymbol{z}}

\newcommand{\hide}[1] {}

\newtheorem{Thm}{Theorem}
\newtheorem{Pro}{Proposition}
\newtheorem{Defn}{Definition}
\newtheorem{Exm}{Example}
\newtheorem{Lem}{Lemma}
\newtheorem{Asm}{Assumption}
\newtheorem{Cor}{Corollary}
\newtheorem{Alg}{Algorithm}
\newtheorem{paRem}{Remark}
\newenvironment{Rem}{\begin{paRem}\rm }{\end{paRem}}

\newcommand{\ColvecTwo}[2]{\left[\ba{c} {#1} \\ {#2} \ea \right]}
\newcommand{\colvecTwo}[2]{\left[\ba{cc} {#1} & {#2} \ea \right]^\top}

\newcommand{\colvecThr}[2]{\left[\ba{ccc} {#1} & \cdots & {#2} \ea \right]^\top}
\newcommand{\ColvecThr}[2]{\left[\ba{c} {#1} \\ \vdots \\ {#2} \ea \right]}
\newcommand{\rowvecThr}[2]{\left[\ba{ccc} {#1} & \cdots & {#2} \ea \right]}
\newcommand{\MatThrTwo}[4]{\left[\ba{ccc} {#1}  & {#3} \\ \vdots & \vdots \\ {#2} & {#4} \ea \right]}

\date{}
\newcommand{\nN}{GN_1}

\title{A sparse recovery method for initial ranging in IEEE 802.16 OFDMA systems}
\author{Md Mashud Hyder and Kaushik Mahata 
\thanks{Department of Electrical Engineering, The University of Newcastle, Australia.}
\thanks{Research is supported by the Australian Research Council.}
}

\begin{document}
\maketitle
\begin{abstract}
Initial ranging constitutes a part of the synchronization procedure 
employed by the wireless communication standards. This allows the 
base station (BS) to detect the subscriber stations (SS) that are willing to 
commence communication. In addition, the ranging process allows the 
BS to estimate the uplink channel parameters of these SSs.
Accurate estimation of these parameters are crucial as they
ensure that the uplink signals from all the SSs arrive at the BS synchronously 
and approximately at the same power level. However, this detection
and estimation problem turns out to be very challenging when multiple users 
initiate the ranging procedure at the same time. We address this issue by exploiting the underlying
sparsity of the estimation problem. We propose a fast sparse signal 
recovery approach to improve the ranging performance in multi-user 
environment. Compared to the standard correlation based 
techniques, our method shows a clear improvement in ranging 
code detection, timing offset and channel power estimation. Although
this method has been developed around the WiMAX standard, 
the underlying principles apply to other OFDM based standards as well.
\end{abstract}


\section{Introduction} 
The orthogonal frequency-division multiple access 
(OFDMA) scheme has been adopted by the IEEE WiMAX standard \cite{rang8}. 
To maintain orthogonality among the subcarriers of 
different users, and to avoid the occurrence of multiple-access 
interference (MAI), the uplink signals arriving at the BS
must be aligned to the local time and frequency 
references. 
For this purpose, the IEEE WiMAX standard enforces
a network entry procedure called initial ranging (IR). 
The ranging process starts with the allocation of a pre-defined set of 
subcarriers by the BS in specific time slots, which is known as a ranging 
opportunity. The SSs which wish to commence communication with the BS,
referred to as the ranging terminals (RTs), can take this opportunity 
by modulating a randomly selected ranging code onto the allocated 
subcarriers. 
Due to different position of RTs within the radio coverage
area, ranging signals transmitted by different RTs arrive at
the BS with their specific transmission time delay. At the
receiving side, the BS is required to extract timing and power
information for each detected code and inform the RTs about the 
extracted information.

Multiuser code separation as
well as their timing and power estimation are the main
tasks of the IR process \cite{rang1}. The correlation based approach 
proposed in \cite{rang9}, is based on the principle that a time delay
can be represented by a phase shift in the frequency domain. 
Lee \cite{rang10} replaces the WiMAX ranging codes by a set of 
generalized chirp-like polyphase sequences to get a more accurate timing 
estimate. The work in \cite{rang2} demonstrates that the frequency-domain 
correlation approach outperforms its time domain counterpart. However, 
the methods \cite{rang9,rang10,rang2} simply treat the MAI as a noise, 
which results in performance degradation in multiuser environment. 
An approach different from the IEEE 802.16 standards has been proposed in \cite{rang11} to overcome the MAI problem in code detection. The idea is to allocate a small number of subcarriers to each ranging opportunity so that most of the RTs are expected to transmit on disjoint sets of subcarriers resulting a minimum level of MAI. However, the reduction of the number of effective subcarrier for each user results in the degradation of timing estimation performance \cite{rang1}.
A similar approach has been proposed in \cite{rang12} for channel synchronization. This method assumes that the uplink signals are transmitted over disjoint subcarriers, and the receivers use filter banks to separate multiuser codes. 
The work in \cite{rang13} improves the ranging performance by dividing 
the ranging signals into several groups with each group being transmitted 
over exclusively assigned subcarriers.
The concept of successive interference cancellation (SIC) has been 
employed in \cite{rang1,rang5}. The method proposed in 
\cite{rang1} works in an iterative fashion where the strongest path of 
each active RT is detected and is removed from the received signal and 
the resulting signal is used in succeeding iterations. The work 
\cite{rang5} differs from \cite{rang1} in the aspect is that in 
\cite{rang5} one additional ranging signal is detected at each iteration 
instead of a single multipath component. 

Sparse signal representation \cite{cs5,dan} has many potential applications, including spectral analysis \cite{jlzadoa}, channel estimation \cite{channel2} etc.
In this paper, we pose the problem of user detection and channel estimation in ranging process as a sparse recovery problem. This approach is founded on the two observations:
\begin{itemize}
\item The number of ranging subscriber stations is much smaller than the number of ranging codes.
\item Only a few channel taps are of some noticeable magnitude, and the remaining vast majority of channel taps are of negligible magnitude.
\end{itemize}
We first develop a signal model that allows us to exploit the above facts and pose the ranging problem as a sparse recovery problem. By applying the standard sparse recover methods to solve the sparse recovery problem arising in ranging problem, we found that associated computation time is too high compared to what is typically allowed by the WiMAX standard. For this reason, we propose a new algorithm to solve the sparse-ranging problem. 
The proposed algorithm combines two different types of sparse recovery 
algorithms to provide a time efficient solution. In particular, we apply a non-convex sparse recovery algorithm which has fast convergence property if it initialized sufficiently close to the final solution. To obtain a good initial input, we consider another convex optimization algorithm that can provide a good estimate of the final solution after a few iterations.  
The proposed handover algorithm 
computes most of its multiplications by using Fast Fourier Transform (FFT). 
Due to noise contribution in the ranging signal, the recovered signal from the algorithm may not be truly sparse. For this reason, we analyse the statistical property of the recovery error. This analysis is used to formulate a systematic hypothesis test to detect the codes and the associated timing offset.

{\em Notation:} Lowercase boldface letters denote vectors. The $k$-th component of a vector $\zx$ is denoted by $[\zx]_k$. Uppercase boldface letters are used to represent matrices. For a matrix $\bA$, we use $[\bA]_{ij}$ to denote the element of $\bA$ at its $i$-th row and $j$-th column. The $\mathbb{Z}$ denotes the set of all integers. We use $(\bA)^*$, and $(\bA)^{\intercal}$ to denote complex conjugate transpose and transpose of the matrix respectively. $\bf{0}_n$ and $\bf{1}_n$ are column vectors of size $n$ with all $0$'s and $1$'s respectively.

\section{Signal Model}
\label{sec:model}

\subsection{Single ranging terminal}

Consider an uplink OFDMA system with $N$ subcarriers. Thus, 
each OFDM symbol contains $N$ data symbols. In addition, an OFDM symbol 
must contain a cyclic prefix. Let the length of the cyclic prefix be 
$N_g$.
Therefore, the length of a OFDM  symbol is
\[
\bar{N} = N + N_g.
\]

Now consider a particular transmitter T in this system.
Let $u(k), \ k \in \mathbb{Z}$ be the 
sequence of the channel symbols transmitted by T. These channel 
symbols are grouped in OFDM symbols. Without any loss of generality
we assume that the $n$th OFDM symbol $\zu_n$ 
consists of $u(k), \ k = n\bar{N}, n\bar{N}+1, \ldots, 
n\bar{N} + \bar{N} - 1$. The $n$-th OFDM symbol vector is written compactly as
\be
\zu_n = [ \ u(n\bar{N}) \ \ u(n\bar{N} + 1) \ \ \cdots \ \
u(n\bar{N} + \bar{N} - 1) \ ]^{\intercal}.
\label{OFDM_FRAME_INDICES}
\ee

Let us denote the impulse response coefficients between the transmitter
T and the base station as $h(p), p = 0,1,\ldots,P-1$. Consequently, the
contribution $v(k), k \in \mathbb{Z}$ of transmitter T in the signal
received by the base station is given by
\be
v(k) = \sum_{p=0}^{N-1} h_p u(k-p -d).
\label{CONVOLUTION}
\ee
In practice, the value of $N$ is more than 200, and typically $h(p) = 0$
for all $p > 50$.  
The delay $d$ depends on the distance between T and BS.
One purpose of the initial ranging process is to estimate 
$d$ so that the transmitter can align its transmission with the frame
boundaries of the base station. In a cellular communication architecture
$d$ cannot be arbitrarily large. 
It must be bounded. In IEEE802.16 the cell radius is chosen such
that $d < N$.

IEEE802.16 identifies some specific uplink subcarriers as `ranging 
subchannels'. In the sequel we assume that there are $M$ subcarriers
in the group of ranging subchannels, and denote their indices by 
$\{j_m : m = 1, 2, \ldots,M\}$. The value of $M$ in IEEE802.16
is 144.
In IEEE802.16 a `ranging opportunity' spans over two consecutive
OFDM symbols when the `ranging terminals', who wish start communicating
via the
base station, can send their ranging codes. The ranging codes must be
sent via the `ranging subchannels'. 

Consider a ranging opportunity
consisting of the OFDM symbols $n-1$ and $n$, respectively. Suppose 
T is a ranging terminal who wants to use this ranging opportunity.
According to IEEE802.16, T must construct the OFDM symbols $\zu_{n-1}$
and $\zu_n$ as follows. First it chooses a column $\zc_{\ell}$ of 
a pre-specified $M \times G$ code matrix
\[
\bC = [ \ \zc_1 \ \ \zc_2 \ \ \cdots \ \ \zc_G \ ]
\]
uniformly at random. Thus the probability that $\ell$ is a given 
integer in the set $\{1,2,\ldots,G\}$ is $1/G$. IEEE802.16 specification
defines $\bC$ preciely. Hence it is known to the BS and all the 
transmitters including T. After choosing $\ell$ the terminal T 
calculates the numbers
\be
s_q = \frac{1}{\sqrt{N}} \sum_{m=1}^M [\bC]_{m,\ell} 
\exp \{ \irm 2 \pi j_m q / N \}, \
q = 0,1,\ldots,N-1,
\label{CALCULATE_S}
\ee
where $[\bC]_{m,\ell}$ denotes the $m$ th component of $\zc_{\ell}$,
which is also the element at the $m$ th row and $\ell$ th column of 
$\bC$. The 
operation \cf{CALCULATE_S} can be seen as the process of modulating 
the $j_m$ th subcarrier by $\bC_{m,\ell}$, and modulating the non-ranging
subcarriers with $0$. In practice this computation is carried out using
the IFFT algorithm, and is compactly given as
\be
\zs := [ \ s_{0} \ \ s_{1} \ \ \cdots \ \ s_{N-1} \ ]^{\intercal}
= \bF^* \GTh^{\intercal} \zc_{\ell},
\label{EXPRESSION_BS}
\ee
where $\bF$ is the  $N \times N$ FFT matrix such that
\[
[ \bF ]_{km} = \exp \{ - 2 \pi \irm  (k-1)(m-1) / N \} / \sqrt{N},  
\psp \irm  = \sqrt{-1}.
\]
The matrix $\GTh$
is an $M \times N$ row selector matrix such that the $m$ th row of $\GTh$
is the $j_m$ th row of the $N \times N$ identity matrix. 
Recall that $j_m, m=1,2,\ldots,M$ are the indices associated with the  
ranging subchannels. 
The process of
modulating the subcarriers via IFFT is equivalent of pre-multiplication
by $\bF^*$ in \cf{EXPRESSION_BS}. 
The premultipaction of $\zc_{\ell}$ by $\GTh^{\intercal}$ implies that only
the ranging subcarriers are modulated by the
appropriate components of the ranging code $\zc_{\ell}$.

Using $\Bs$ the transmitter constructs the OFDM
symbols $\zu_{n-1}$ and $\zu_n$ as
\begin{align}
\nonumber
\zu_{n-1}
&= 
[ \ 
s_{N-N_g} \ \
\cdots \ \
s_{N-2} \ \
s_{N-1} \ \
s_0 \ \
s_1 \ \
\cdots \ \
s_{N-1} \ ]^{\intercal},
\\
\zu_{n}
&=
[ \
s_0 \ \
s_1 \ \
\cdots \ \
s_{N-1} \ \
s_0 \ \
s_1 \ \
\cdots \ \
s_{N_g-1} 
\ ]^{\intercal}.
\label{RANGING_FRAMES}
\end{align}

To detect the ranging codes
the base station works with the first $N$ samples of the $n$th received 
OFDM symbol. According to the standard practice in OFDMA, the BS computes
an FFT of these samples and then examines the data
received in the ranging subchannels.

We first find the contribution 
\[
\zv_n = [ \ v(n\bar{N}) \ \ v(n\bar{N} + 1) \ \ \cdots \ \
v(n\bar{N} + N - 1) \ ]^{\intercal}
\]
of T in the first $N$ samples of the
$n$th OFDM symbol received by the base station in terms of $\Bs$
and
\be
\zh = [ \ h_0 \ \ h_1 \ \ \cdots \ \ h_{N-1} \ ]^{\intercal}.
\label{eq:channel}
\ee
Using \cf{OFDM_FRAME_INDICES} and \cf{RANGING_FRAMES} note that
both $\zu_{n-1}$ and $\zu_n$ are linear functions of $\Bs$.
In particular,
\be
u(n \bar{N} - k) = \left \{ \ba{lc}
s(k-N)       & -\bar{N} < k \le -N, \\ 
s(k)         & -N < k \le 0, \\
s(N-k),  &  0 < k \le N, \\
s(2N-k), &  N < k \le \bar{N}.
\ea \right.
\label{U_AND_S}
\ee
In the following we denote the circular shift operator by 
$\downarrow(\cdot)$. For instance, the circularly shifted version of 
$\zs$ by $k$ places is given as 
\be
\zs_{\downarrow(k)} 
:=
[ \ s_{N-k} \ \ \cdots \ \ s_{N-1} \ \ s_0
\ \ s_1 \ \ \cdots \ \ s_{N-k-1} \ ]^{\intercal}.
\label{CIRCULAR-SHIFT}
\ee
A few steps of algebra using \cf{CONVOLUTION} and 
\cf{U_AND_S}, and using the fact that $d < N$, we get 
\be
\zv_n = 
\bH \zs_{\downarrow(d)}, 
\label{EXPRESSION_ZV}
\ee
where 
$\bH$ is an $N \times N$ cyclic matrix
\be
\bH = [ \ \zh_{\downarrow(0)} \ \ \zh_{\downarrow(1)} \ \
\cdots \ \ \zh_{\downarrow(N-1)} \ ].
\label{CYCLIC_H}
\ee
Expressions \cf{EXPRESSION_ZV} and \cf{CYCLIC_H} involving circularly
shifted versions of $\Bs$ and $\Bh$ are typical of OFDM, and are
resulted from the way the OFDM symbols $\zu_{n}$ and $\zu_{n-1}$ are
constructed in \cf{RANGING_FRAMES}. This special construction allows
us to exploit the identity \cite{oppen} 
\begin{align}
\nonumber
\bF \zx_{\downarrow(k)} 
&= \mathrm{diag}(\bF(:,k+1)) \bF \zx \\
&= \mathrm{diag}(\bF \zx) \bF(:,k+1),
\ \ k=0,1,\ldots,N-1,
\label{FFT_IDENTITY}
\end{align}
satisfied by the FFT matrix $\bF$, and any vector $\zx$. 
Note that we use the standard MATLAB notation $\bF(:,k)$ to denote 
the $k$th column of $\bF$, and $\mathrm{diag}(\zx)$ to denote the
diagonal matrix such that
$[\mathrm{diag}(\zx)]_{k,k} = [\zx]_k$. 
Recall
that for detecting the ranging codes the BS must compute
an FFT of $\zv_n$, and then extract the data
received in the ranging subchannels. The FFT of $\zv_n$ is $\bF \zv_n$.
Hence, the data received from T in the ranging subchannels is given
by premultiplying $\bF \zv_n$ by the row selector matrix $\GTh$.
Using \cf{CYCLIC_H} and \cf{FFT_IDENTITY} it follows that
\[
\bF \bH = 
\mathrm{diag}(\bF \zh) \bF.
\]
Note that $\hat{\zh} := \bF \zh$ is the FFT of $\zh$. Hence by 
\cf{EXPRESSION_ZV} and \cf{FFT_IDENTITY} if follows that
\begin{align}
\nonumber
\bF \zv_n 
&= 
\bF \bH \zs_{\downarrow(d)} 
=
\mathrm{diag}(\hat{\zh}) \ \bF  \zs_{\downarrow(d)}
\\
\nonumber
&=
\mathrm{diag}(\hat{\zh}) \ \mathrm{diag}(\bF(:,d+1)) \ \bF \zs \\
\nonumber
&=
\mathrm{diag}(\bF \zs) \ \mathrm{diag}(\bF(:,d+1)) \ \hat{\zh} \\
&=
\mathrm{diag}(\bF \zs) \bF \zh_{\downarrow(d)}.
\end{align}
It is well-known that $\bF$ is an unitary matrix, i.e., $\bF^* \bF
= \bF \bF^* = \bI$. Hence \cf{EXPRESSION_BS} implies $\bF \zs = 
\GTh^{\intercal} \zc_{\ell}$. Hence the data received by the BS at the 
ranging subchannels due to transmission of T is given by
\be
\GTh \bF \zv_n = \bE_{\ell} \zh_{\downarrow(d)},
\psp
\bE_{\ell} =
\GTh \
\mathrm{diag}( \GTh^{\intercal}  \zc_{\ell} ) \
\bF.
\label{ONE_TERMINAL_RELATION}
\ee
Each of the matrices $\bE_{\ell}, \ell = 1,2,\ldots,G$ is of size
$M \times N$, and is known because $\Bc_{\ell}$ is known.

Typically, we  know a number $P$ such that
$|h_k| = 0$ for $k \ge P$. At this point we emphasize that OFDM can 
effectively equalize the inter-symbol interference effects
only when 
$P < N_g$. Thus the existence of the upper bound $P$ is a key assumption
in OFDM. In addition, the cell radius gives an 
upper bound $D$ on $d$. Hence $d+P < D+N_g$. By construction of 
$\zh_{\downarrow(d)}$, we know only first $D+P$ of its 
rows are non-zero. Hence it is fine to truncate $\zh_{\downarrow(d)}$
to a $D+N_g$ dimensional vector, and thus it is enough to 
work with only first $D+N_g$ columns of $\bE_{\ell}$.

\subsection{Multiple ranging terminals}

So far we have considered a single ranging terminal, and 
in \cf{ONE_TERMINAL_RELATION} we have 
quantified its contribution to the data vector received by the BS in 
the ranging subchannels. In this section, we generalize the analysis 
for multiple ranging terminals, and account of the receiver noise
at the base station.

Suppose that the code $\zc_{\ell}$ is chosen and 
transmitted by $\tilde{N}_{\ell}$
number of ranging terminals. This means that the total number ranging
terminals in the system is 
$K = \tilde{N}_1 + \tilde{N}_2 + \cdots +
\tilde{N}_G$. We emphasize that $\tilde{N}_{\ell}$ is a random quantity
for a given $\ell$. Typically $K < 10$, $G=256$, 
and the probability that $\tilde{N}_{\ell} > 1$ is 
\[
\mathrm{Prob} \{ \tilde{N}_{\ell} > 1 \}
=
1 
-
(1-1/G)^{K} 
- 
K (1-1/G)^{K-1}(1/G),
\]
which is a very small number. Hence the probability that two or more
ranging terminals will collide by selecting the same code is very 
small. However for the sake of generality we do not exclude that 
possibility. 

Let $\zh_{\ell}^{(k)}$ and $d_{\ell,k}, \ \ k=1,2,\ldots,
\tilde{N}_{\ell}$ denote the channel impulse response vector and
the delay of the $k$th ranging terminal transmitting the code 
$\zc_{\ell}$. Then by the principle of superposition the data vector
$\By$
received by the BS at the ranging subchannels is given by,
see \cf{ONE_TERMINAL_RELATION}
\be
\zy = \sum_{\ell = 1}^G \bE_{\ell} \zh_{\ell} + \ze,
\label{MULTI_TERMINAL_RELATIONSHIP}
\ee
where $\ze$ is the additive receiver noise, and 
\be
\label{eq:hl}
\zh_{\ell} = 
\left\{
\ba{cl}
\sum_{k=1}^{\tilde{N}_{\ell}} 
[ \zh_{\ell}^{(k)} ]_{\downarrow(d_{\ell,k})}, & \tilde{N}_{\ell} > 0,\\
0, & \tilde{N}_{\ell} = 0.
\ea
\right.
\ee
is the combined channel vector for all the ranging terminals transmitting
the code $\zc_{\ell}$. Note that the power received by the BS 
corresponding to the ranging code $\zc_{\ell}$ is given by 
\cite{rang1}:
$$
\Gamma_{\ell}= \zh_{\ell}^* \zh_{\ell}.
$$

\section{Estimation of ranging information}

\subsection{Formal problem Statement}
\label{sec:pro}

Given $\zy$ the signal model in \eqref{MULTI_TERMINAL_RELATIONSHIP} the BS needs to
\begin{enumerate}
\item Find the set $\mathcal{L} = \{ \ell : \Gamma_{\ell} \ne 0 \}$;
\item For every $\ell \in \mathcal{L}$ find $\Gamma_{\ell}$ and 
$d_{\ell,1}$ assuming $\tilde{N}_{\ell} = 1, \ \forall \ell \in \mathcal{L}$.  
\end{enumerate}
In some rare cases $\tilde{N}_{\ell} > 1$ for some $\ell$. In this case 
the ranging requests of the users who chose $\Bc_{\ell}$ simultaneously
would collide. Nevertheless, as we see later, the BS would detect that $\Bc_{\ell}$ 
was transmitted among others. A IEEE802.16 base station allocates some
small bandwidth corresponding to every detected ranging code. The 
ranging terminals use  this `grant' to transmit their buffer status 
information, and expect to obtain some adequate amount of bandwidth from 
the BS to commence data transmission. However, if two ranging terminals,
say T1 and T2, chose the same code $\Bc_{\ell}$ for ranging request, 
then the estimate of $\Gamma_{\ell}$ and $d_{\ell,1}$ obtained by the BS
during the ranging process would have no physical meaning. In addition,
the bandwidth request from T1 and T2 will collide again. Consequently,
the BS will not be able to decode the bandwidth request data from T1 and 
T2. In such a scenario a IEEE802.16 BS does not allocate any further 
bandwidth corresponding to code $\Bc_{\ell}$, and after a timeout 
period T1 and T2 commence the ranging process again \cite{rang9,rang1}.

Recall that the first $d$ components of
$\Bh_{\downarrow(d)}$ are zero, see \cf{CIRCULAR-SHIFT}. 
Hence by construction of $\zh_{\ell}$ in \eqref{eq:hl}, the index of 
the first nonzero component of $\zh_{\ell}$ is $1 + \bar{d}_{\ell}$, 
where
\[
\bar{d}_{\ell} = \min_{k \in \{ 1,2,\ldots,\tilde{N}_{\ell} \} }
\
d_{\ell,k},
\psp \ell \in \mathcal{L}.
\]
Clearly, if $\tilde{N}_{\ell} = 1$, then $\bar{d}_{\ell} = d_{\ell,1}$.
For this reason we propose to estimate $\bar{d}_{\ell}$ as the timing 
offset corresponding to an $\ell \in \mathcal{L}$. When 
$\tilde{N}_{\ell} = 1$ this estimate is consistent with our requirements.
On the other hand, if $\tilde{N}_{\ell} > 1$, this estimate will have no
practical relevance, for, as discussed above, BS will reject $\Bc_{\ell}$
in the bandwidth request stage.

\subsection{Sparse recovery framework}\label{sec:srf}

Recall that for any $\ell$ 
only first 
\begin{align}\label{eq:el}
N_1 := D + N_g
\end{align}
components of $\zh_{\ell}$ are non-zero. Hence
\[
\bE_{\ell} \zh_{\ell} = \bE_{\ell}(:,N_1) \ \zh_{\ell} (1:N_1).
\]
Note that we use Matlab notation $\bE_{\ell}(:,1:N_1)$ to denote
the submatrix of $\bE_{\ell}$ formed by taking its first $N_1$
columns. Similarly, $\zh_{\ell} (1:N_1)$ denotes the vector formed
by taking the first $N_1$ components of $\zh_{\ell}$. 
Then we can write \cf{MULTI_TERMINAL_RELATIONSHIP} as
\be
\zy = \bA \zx + \Be,
\label{MEASUREMENT_EQUATION}
\ee
where 
\begin{align}
\nonumber
\zx &:= [ \ \zh_1^{\intercal} (1:N_1) \ \ 
\zh_2^{\intercal} (1:N_1) \ \ \cdots \ \ 
\zh_{G}^{\intercal} (1:N_1) \ ]^{\intercal},
\\
\nonumber
\bA &= [ \ \bE_1 (:,1:N_1) \ \ \bE_2 (:,1:N_1)
\ \ \cdots \ \ \bE_G(:,1:N_1) \ ].
\end{align}
Note that by definition of $\bE_{\ell}$ in \cf{ONE_TERMINAL_RELATION}, 
$\bA$ is a known matrix. On the other hand $\zx$ and $\Be$ are 
unknowns.
Typically, the total number of ranging terminals $K = \sum_{\ell = 1}^G
\tilde{N}_{\ell} \ll G$, implying $\tilde{N}_{\ell} = 0$ 
(and therefore $\zh_{\ell} = 0$) for a vast 
majority of the values $\ell \in \{1,2,\ldots,G\}$. This makes $\Bx$
very sparse. This observation motivates a sparse recovery framework for
solving the ranging problem.

We propose to estimate a sparse vector $\zx$ that is consistent with
\cf{MEASUREMENT_EQUATION}. There are many reliable algorithms for
solving such sparse estimation problems \cite{cs5,dan}. Denote the sparse estimate by $\breve{\zx}$.
Then the BS can extract the required ranging information as follows. 
Partition $\breve{\zx}$ into $G$ number of sub-vectors:
\[
\breve{\zx} = [ \ \breve{\zh}_1^{\intercal} \ \ \breve{\zh}_2^{\intercal} \ \ 
\cdots \ \ \breve{\zh}_G^{\intercal} \ ]^{\intercal},
\] 
where each $\breve{\zh}_{\ell}$ is of length $N_1$. 
Then we declare $\ell \in \mathcal{L}$ only
if $\|\breve{\zh}_{\ell}\| \ne 0$ and the index of the 
first nonzero component of $\breve{\zh}_{\ell}$ leads to an estimate of 
$\bar{d}_{\ell}$.
 
\subsection{Background on sparse recovery methods} 
 
If $\Be = 0$, then the ideal way to reconstruct a sparse $\zx$ from 
$\zy$ requires solving 
\begin{align}
\label{eq:l0}  
\zx_*= \arg \min_{\zv} \| \zv \| _0 \ \ \ \mathrm{subject} \
\mathrm{to} \ \ \ \zy=\bA \zv,
\end{align}
where $\| \zv \|_0$, which denotes the $\ell_0$ norm of a vector $\zv$, is simply the number of non-zero components in $\zv$. Thus the idea is
to find $\zx_*$ with the smallest number of non-zero components 
satisfying $\By = \bA \zx_*$. The unique representation theorem 
\cite{foc1} ensures that under mild technical conditions 
there is a unique $\zx_*$ with $|| \Bx_* ||_0
< M/2$ satisfying  $\zy = \bA \zx_*$.

However, \cf{eq:l0} is combinatorial in nature \cite{ad2}. 
The most popular alternative approach for relaxing \eqref{eq:l0} is 
called Basis Pursuit (BP) \cite{ad1, cs4}, where the $\ell_0$ norm in  
\eqref{eq:l0} is replaced by $\ell_1$ norm: 
\begin{align}
\label{eq:l1}  
\zx_* = \arg \min_{\zv} \| \zv \| _1 
\ \ \ \mathrm{subject} \
\mathrm{to} \ \ \ \zy=\bA \zv.
\end{align}
Here
\[
||\zv||_1 := \sum_{k=1}^{\nN} | [ \zv ]_k |.
\]
BP can be posed as a linear program \cite{ad1} over second order
cones, and can be solved
in polynomial time. In addition, it has been shown in 
\cite{ad1} that BP recovers the sparsest solution 
to $\zy = \bA \zx$ with a very high probability.

The above simple ideas can be adapted quite well even when $\Be \ne 0$
\cite{cs5}. However, the existing
algorithms for solving \cf{MEASUREMENT_EQUATION} are unable to 
converge to a satisfactory solution within the time-frame 
available to solve the ranging problem in practice. In the sequel
we propose a new approach to overcome this hurdle. 

Our approach blends the nice properties of the $\ell_0$ and $\ell_1$
methods. The so called $\ell_0$ approximation methods \cite{l06,isl0,
ejlza}
are known to converge very fast if initialized sufficiently close to 
the final solution. But these methods being non-convex, may often 
get trapped in some local optimal point when initialized far away from 
the final solution. The $\ell_1$ methods being
convex, does not have the local optima problem, but typically take 
a large number of iterations for convergence. Therefore, we aim to 
to start with an $\ell_1$ approach and then handover to an $\ell_0$
approach when the solution is `sufficiently close'.
In particular, we  introduce a new $\ell_1$ norm minimization 
method that can obtain a rough estimate of $\zx$ in 
only a few iterations, and then handover to a reliable
$\ell_0$-approximation algorithm \cite{isl0,ejlza}.

\subsection{$\ell_1$ optimization algorithm}

In this paper we propose to solve a special dual of \cf{eq:l1}. 
This
dual formulation relies on the theory of minimum norm problems in 
Banach spaces. 
Given element $\zu \in \mathbb{C}^{\nN}$ 
we define the infinite norm  as
\[
|| \zu ||_{\infty} = \max_{k \in \{ 1,2,\ldots, \nN \} } 
\ | [ \zu ]_k |.
\]
Let us define the bilinear from $\langle \cdot , \cdot \rangle$ as
\[
\langle \zv, \zu \rangle := \frac{1}{2} ( \zu^* \zv + \zv^* \zu ).
\]
Note that this bilinear form maps 
$\mathbb{C}^{\nN} \times \mathbb{C}^{\nN}$ onto 
$\mathbb{R}$. By H{\" o}lder's inequality it follows that 
\be
\langle \zv, \zu \rangle \le ||\zv||_1 ||\zu||_{\infty},
\label{HOLDER}
\ee
provided that both $||\zv||_1$ and 
$||\zu||_{\infty}$ exist. In addition, when we have an equality 
in \cf{HOLDER}, then we say $\zv$ and $\zu$ are aligned.
The condition
for alignment can be verified to be as follows \cite{Luenberge}: 
\begin{Pro}
Let
\[
\mathcal{K} = \{ k : | [\zu]_k | = || \zu ||_{\infty} \}.
\] 
Then $\langle \zv, \zu \rangle =
||\zv||_1 ||\zu||_{\infty}$ only if
\bit
\item $[\zv]_k = 0$, for all $k \notin \mathcal{K}$;
\item For all $k \in \mathcal{K}$ it holds that $[\zv]_k = 
\mu_k \mathrm{conj}( [ \zu]_k ) / | [ \zu]_k |$ 
for some non-negative number $\mu_k \in \mathbb{R}$.
\eit 
\end{Pro}
We are now ready to state the key result allowing us to formulate
a convenient dual of \cf{eq:l1}.

\begin{Thm} Let us define the sets
\[
\mathbb{V} = \{ \zv: \bA \zv = \zy \},
\psp
\mathbb{U} = \{ \zg: || \bA^* \zg ||_{\infty} \le 1 \}.
\]
Then 
\be
\min_{\zv \in \mathbb{V}} || \zv ||_1
=
\max_{\zg \in \mathbb{U}} \frac{1}{2} ( \zy^* \zg + \zg^* \zy ).
\label{DUAL1}
\ee
In addition, let $\zx_*$ be the solution to the optimization 
problem in the left hand side of \eqref{DUAL1}, and let $\zg_*$
be the solution to the optimization 
problem in the right hand side of \eqref{DUAL1}. Then 
\be
\langle \zx_*, \bA^* \zg_* \rangle = || \zx_*||_1 \
|| \bA^* \zg_*||_{\infty}.
\label{ALIGNMENT}
\ee
\label{THM_1}
\end{Thm}
{\em Proof:} See \cite{Luenberge}. \feop
Note that $\Bg$ is of significantly smaller size compared to $\Bx$.
Computationally it is a lot more economical to solve the dual in 
the right hand side of \cf{DUAL1}, and apply the alignment condition
\cf{ALIGNMENT} to recover $\Bx_*$ from $\Bg_*$. We write the dual
problem as 
\begin{align}
\label{eq:prim}
\zg_*
=
\arg \max_{\zg} \ & \frac{1}{2}(\zg^*\zy+\zy^*\zg)
\\
\label{eq:pcon} 
\mathrm{subject~ to} \ &
\zg^* \za_i\za_i^*\zg \le 1, \  \ \ 
i=1,\ldots \nN
\end{align}
where the $i$ th column of $\bA$ is denoted by $\za_i$. 
We wish to solve it via a primal-dual algorithm \cite{co}. 
Therefore we write the Lagrangian associated with 
\cf{eq:prim}-\cf{eq:pcon}:
\be
L(\zg,\zz) = \frac{1}{2} 
\{ \zg^*\zy+\zy^*\zg - \zg^* \bA 
\mathrm{diag}(\zz) \bA^* \zg + {\bf 1}^{\intercal}  \zz \}
\ee
where ${\bf 1}$ is a $\nN$ dimensional vector of all ones, and
$\zz/2$ is the real-valued vector of Largange multipliers. 
Each component of $\zz$ must be non-negative, and we denote it
by $\zz \ge 0$. Now it is a standard result in the theory of least squares that
\begin{align}
\arg \max_{\zg} L(\zg,\zz) &= 
[\bA \ \mathrm{diag}(\zz) \bA^*]^{-1} \zy, 
\label{ARGMAX_G}
\\
\max_{\zg} L(\zg,\zz) &= \frac{1}{2} \{ {\bf 1}^{\intercal} \zz
+ \zy^* [\bA \  \mathrm{diag}(\zz) \bA^*]^{-1} \zy \}.
\end{align}
Therefore, we can obtain $\zz_*$ by solving the Lagrangian dual
of \cf{eq:prim}-\cf{eq:pcon}:
\begin{align}
\nonumber
\zz_*
=
\arg \max_{\zz} \ & 
\frac{1}{2} \{ {\bf 1}^{\intercal} \zz
+ \zy^* [\bA \ \mathrm{diag}(\zz) \bA^*]^{-1} \zy \}
\\
\mathrm{subject~ to} \ &
\zz \ge 0.
\label{DUAL2} 
\end{align}
In reality, a primal-dual algorithm would solve 
\cf{eq:prim}-\cf{eq:pcon} and its Lagrangian dual \cf{DUAL2}
together by finding the solution to the Karush-Kuhn-Tucker (KKT)
conditions
\begin{align}
\label{kkt1}
&
\zy = \bA \ \mathrm{diag}(\zz_*) \ \bA^* \zg_{*},
\\
\label{kkt2}
&
[\zz_*]_i \ge 0, \ \ i=1,\ldots,\nN,
\\
\label{kkt3}
&
(1 - \zg_*^* \za_i \za_i^* \zg_*) \ge 0, \ \ \ 
 i=1,\ldots,\nN,
\\
\label{kkt4}
&
[\zz_*]_i \ (1 - \zg_*^* \za_i \za_i^* \zg_*) = 0,
\ \ i=1,\ldots,\nN.
\end{align} 
Equation \cf{kkt1} follows from \cf{ARGMAX_G}. The inequalities  
\cf{kkt2} and \cf{kkt3} must hold because the constraints 
in \cf{eq:pcon} and \cf{DUAL2} must hold. Equation \cf{kkt4}
is the complementary slackness condition which says that for 
any $i$ either $(1 - \zg_*^* \za_i \za_i^* \zg_*) = 0$ or 
$[ \zz]_i=0$. These relations can be used to calculate $\zx_*$
from $\zz_*$ and $\zg_*$. 

\begin{Pro} The optimal solution $\zx_*$ of \cf{eq:l1} is 
given in terms of $\zz_*$ and $\zg_*$ as
\be
\zx_* = \mathrm{diag}(\zz_*) \ \bA^* \zg_{*}.
\label{XGZ}
\ee
\end{Pro}

{\em Proof:} Note that by setting $\zx_*$ as in \cf{XGZ} we 
do satisfy $\zy = \bA \zx_*$. It remains to verify the alignment
condition \cf{ALIGNMENT}. 

Now, it must hold that $|| \bA^* \zg_* ||_{\infty} = 1$. This is because
if $|| \bA^* \zg_* ||_{\infty} < 1$, then we could always 
multiply $\zg_*$ by a suitable real valued 
scalar $\kappa > 1$ such that 
$|| \bA^* (\kappa \zg_*) ||_{\infty} = 1$, and  
\[
\zg_*^* \zy + \zy^* \zg_* < (\kappa \zg_*)^* \zy + \zy^* 
(\kappa \zg_*),
\]
leading to a contradiction. 

Let us define
\[
\mathcal{K} = \{ i: \zg_*^* \za_i \za_i^* \zg_* = 1 \}.
\]
Note that $\mathcal{K}$ is nonempty since 
$|| \bA^* \zg_* ||_{\infty} = 1$. However, the complementary slackness
condition \cf{kkt4} implies that $[\zz_*]_i = 0$ for all 
$i \notin \mathcal{K}$. Hence
\be
|| \zx_* ||_1 = \sum_{i \in \mathcal{K}} 
[\zz_*]_i |\za_i^* \zg_*| = \sum_{i \in \mathcal{K}} 
[\zz_*]_i.
\ee
The last equality follows because by definition of $\mathcal{K}$
we have $|\za_i^* \zg| = 1$ for all $i \in \mathcal{K}$. Then
\[
\zg_*^* \bA \zx_* = \sum_{i \in \mathcal{K}}
\zg_*^* \za_i \za^*_i \zg [\zz_*]_i = 
\sum_{i \in \mathcal{K}} 
[\zz_*]_i = || \zx_* ||_1
\]
Hence we can verify that the alignment condition
\bear
\langle \zx_*, \bA^* \zg_* \rangle
&=& 
\frac{1}{2} ( \zg_*^* \bA \zx_* + \zx_*^* \bA^* \zg_* ) 
=
1 \times || \zx_*||_1  \\
&=& || \bA^* \zg_* ||_{\infty} ||\zx_*||_1
\eear
holds, and thereby the proof is complete.
\feop

We wish to find a numerical method to solve the KKT equations \eqref{kkt1}-\eqref{kkt4},
which are nonlinear simultaneous equations in $\zz_*$ and $\zg_*$. 
To solve the KKT equations using the primal-dual method, one can relax the complementary slackness condition in \eqref{kkt4} to
\begin{align}\label{kkt5}
[\zz_*]_i \ (1 - \zg_*^* \za_i \za_i^* \zg_*) = \mu,
\ \ i=1,\ldots,\nN.
\end{align}
where $\mu>0$. The value of $\mu$ decreases as we progress through the iterations of the primal-dual algorithm. 
The standard way to handle the modified KKT equations \eqref{kkt1}-\eqref{kkt3} and \eqref{kkt5} is to use the Newton's approach.
A derivation of the primal-dual algorithm can be found in \cite{co}, and in our case it reduces to the form summarized in Table-\ref{tab:algo2}, where we define the function
\begin{align}
f_i(\zg)=\zg^* \za_i\za_i^*\zg-1, \ \mathrm{for} \ \ i=1,2,\cdots GN_1
\end{align}
with $f(\zg)=[f_1(\zg) \ f_2(\zg) \cdots f_{GN_1}(\zg)]^{\intercal}$ and vectors $\zq$ and $\zb$ such that
\begin{align}
\zq&=\bA^*\zg\nonumber\\
[\zb]_i&=\frac{1}{f_i(\zg)}, \  \mathrm{for} \ \ i=1,\ldots,\nN.
\end{align}
The Jacobean matrix of $f(\zg)$ turns out to be
\begin{align}
\bJ=
\left[ 
\za_1\za_1^*\zg \ \ \za_2\za_2^*\zg \cdots \za_{GN_1}\za_{GN_1}^*\zg
 \right]^{\intercal}=\mathrm{diag}(\zq)^*\bA^*.
\end{align}
The algorithm in Table-\ref{tab:algo2} terminates when a rough estimation of $\zx$ has been obtained. We use the following procedure at every iteration to check whether the algorithm has yield a sparse enough estimate of $\zx$. Note that $\zx$ is sparse and hence most of its energy should be concentrated on a few number of its components. In fact, it has been shown in \cite{foc1} that any sparse recovery algorithm can perform well when the energy of $\zx$ concentrates within $M/2$ number of its components. At Step-5 in Table-\ref{tab:algo2}, we compute $\hat{\zx}$. Let $\grave{\zx}$ be the thresholded vector constructed from $\hat{\zx}$ by retaining $M/2$ most significant components of $\hat{\zx}$ while setting others to zero. Now compute 
$$\kappa=\frac{\|\grave{\zx}\|_2^2}{\|\hat{\zx}\|_2^2}.$$
The value of $\kappa$ can be used as an indicator to measure the proximity of $\hat{\zx}$ to the actual $\zx$, while $\kappa\rightarrow 1$ indicates that $\hat{\zx}$ may be very close to $\zx$. However, the algorithm in Table-\ref{tab:algo2} targets to produce only a rough estimate of $\zx$ and hence we do not wait until $\kappa=1$. In particular, the termination value of $\kappa$ trades-off computational complexity with estimation accuracy. A larger terminating $\kappa$ brings $\hat{\zx}$ closer to $\zx$ for increased number of iterations. The value of $\kappa$ is being an interesting design parameter. We have observed that the primal-dual algorithm can achieve a $\kappa\ge 0.6$ in only $4-5$ iterations. The reason behind taking small number of iterations may be due to the formulation of the primal direction search step i.e., $\Delta \zg$ (see Step 2, Table-\ref{tab:algo2}). The popular non-convex iterative re-weighted least square (IRLS) based algorithms like FOCUSS \cite{foc1}, ISL0 \cite{isl0} use similar formulations to update the estimate of actual sparse signal ($\zx$) in every iterate. It is well known that if IRLS algorithms can avoid local minima then they provide close estimate of actual signal in a small number of iterations. 
\begin{table}[!t]
\renewcommand{\arraystretch}{1.3}
\centering \caption{Primal-dual algorithm}\label{tab:algo2}
\begin{tabular}{l}
 \hline
{\bf Initialization}\\
\ \ \ 1. Set $\zg=\bf{0}, \zz=\bf{0}$, and parameters $\mu,0<\alpha\le 1$.\\
{\bf repeat}\\
\ \ \ 2. Compute primal-dual search directions:\\
\ \ \ \ \ \ $\Delta \zg=(A \ \mathrm{diag}(\zz)A^*-\bJ^*\bS\bJ)^{-1}(-\zy+\mu\bJ^*\zb)$\\
\ \ \ \ \ \ $\Delta \zz=-(\zz+\mu\zb+\bS\bJ\Delta \zg)$\\
\ \ \ \ \ \ where, $\bS=\mathrm{diag}(\mathrm{diag}(f(\zg))^{-1}\zz)$.\\
\ \ \ 3. Find $0<s\le 1$ such that:\\
\ \ \ \ \ \ 3a. $f_i(\zg+s\Delta \zg)\le0,\zz(i)+s\Delta \zz(i)\ge 0;\forall i.$\\
\ \ \ \ \ \ 3b. The norm of residuals has decreased sufficiently:\\
\ \ \ \ \ \ \ \ \ $\|\tau_{\mu}(\zg+s\Delta \zg,\zz+s\Delta\zz)\|_2\le (1-\alpha s)\|\tau_{\mu}(\zg,\zz)\|_2.$\\
\ \ \ 4. Set $ \zg=\zg+s\Delta\zg,\zz=\zz+s\Delta\zz$.\\
\ \ \ 5. Compute $\hat{\zx}=  \mathrm{diag}(\zz)\bA^*\zg$.\\
\ \ \ 6. Set $\mu=\alpha\mu$.\\
{\bf until} \  ({\em A rough estimation of $\zx$ has not been obtained})\\
\hline
\end{tabular}
\end{table}


\subsection{Smoothed $\ell_0$ minimization \cite{isl0}}\label{sec:isl0}
We use the rough estimate of $\zx$ obtained by the primal-dual method to initialize the improved smoothed $\ell_0$ (ISL0) algorithm \cite{isl0}. The ISL0 algorithm is described below. Define the Gaussian functions,
\begin{align}
\label{eq:f} f_\sigma(\alpha)=\mathrm{e}^{-\frac{\alpha^2}{2\sigma^2} }.
\end{align}
Then it is readily verified~\cite{isl0} that, as $\sigma \rightarrow 0$, the function
\begin{align}\label{eq:Fs}
F_\sigma(\zx) = \sum_{t=1}^{GN_1} f_\sigma([\zx]_t)
\end{align}
behaves like $GN_1-\|\zx\|_0$, motivating the following approximate reformulation of  \eqref{eq:l0}:
\begin{equation}
\label{eq:opt} \bar{\zx}_*(\sigma) := \arg \min_{v} -F_\sigma(v), \ \ \ \mathrm{subject \ to} \ \zy=\bA v,
\end{equation}
while taking $\sigma \rightarrow 0$. Like $\|v\|_0$, the function $F_\sigma(v)$ has many local minima for a small $\sigma$. 
 Hence, one solves  \cf{eq:opt} for a large $\sigma$ initially, and successively decrease $\sigma$ using a small factor and solve \cf{eq:opt} repetitively. Finally solves \cf{eq:opt} for $\sigma = \sigma_0$, where $\sigma_0$ is a small
positive number. The work in \cite{isl0} proposed a systematic way to choose $\sigma_0$.
In presence of noise, i.e. when $\ze\not=0$ in \eqref{MEASUREMENT_EQUATION}, the following optimization for ISL0 has been considered in \cite{jlzadoa}:
\begin{align}\label{eq:opt2} 
\zx_*(\sigma) &= \arg \min_{v} \ \ L_{\sigma}(v), \\
\nonumber
L_{\sigma}(v) &:= \ -F_\sigma(v)
+ \frac{ \lambda}{2} ||\zy -
\bA v||_2^2, 
\end{align}
where $\lambda>0$ depends on noise level. A Gauss-Newton type convex-concave procedure is used in \cite{jlzadoa} to minimize $L_{\sigma}$ for a fixed $\sigma$. A detailed description of convergence properties of \cf{eq:opt} and \eqref{eq:opt2} can be found in \cite{isl0,jlzadoa}.  In particular, the following Lemma gives a direction to minimize $L_{\sigma}$ for a fixed $\sigma$.

\begin{Lem}\label{lem1}\cite{jlzadoa}
Define the mapping $\zeta: \mathbb{C}^{GN_1} \rightarrow
\mathbb{C}^{GN_1}$ such that 
\be 
\zeta(\tilde{\zx}) = \lambda 
\left[W_{\sigma}(\tilde{\zx})/\sigma^2 +\lambda \bA^* \bA \right]^{-1} 
\bA^* \zy, 
\label{8}
\ee 
where $W_{\sigma}(\tilde{\zx})$ is a diagonal matrix: \be W_{\sigma}(\tilde{\zx}) = \left[ \ba{ccc}
f_\sigma([\tilde{\zx}]_1) & \cdots & 0 \\ \vdots & \ddots & \vdots \\ 0
& \cdots & f_\sigma([\tilde{\zx}]_{GN_1}) \ea \right]. \label{6} \ee Then
$\zx_*(\sigma) = \zeta \{ \zx_*(\sigma) \}$. In addition, for any $\zu$
there exits a real-valued scalar $\xi \ge 0$ such that \be
L_{\sigma} \{ \kappa \zeta(\zu) + (1-\xi) \zu  \}  \leq
L_{\sigma}(\zu). \label{4} \ee
\end{Lem}
Lemma \ref{lem1} reveals the fact that $\zx_*(\sigma) = \zeta \{ \zx_*(\sigma) \}$, and motivates a fixed-point iteration approach to find $\zx_*(\sigma)$ by solving the equation $\tilde{\zx}= \zeta \{\tilde{\zx} \}$. Furthermore, the Lemma provides a direction such that $L_{\sigma}(\tilde{\zx}) $ is
decreasing along $\zeta(\tilde{\zx})-\tilde{\zx}$.

The ISL0 algorithm is given in Table-\ref{tab:algo}.
\begin{table}[!t]
\renewcommand{\arraystretch}{1.3}
\centering \caption{ISL0 Algorithm}\label{tab:algo}
\begin{tabular}{l}
 \hline
 {\bf Input:} $\tilde{\zx}^{(0)},\sigma_{st}$\\
{\bf Initialization}\\
\ \ \ 1. Set $\sigma=\sigma_{st}$, $\lambda \in [1,100]$ and $\rho, \eta, \gamma \in [0,1) $, \\
\ \ \ \ \ \ $i=0, \sigma_0 \in [0.1,10^{-4}]$.\\
{\bf repeat}\\
\ \ \ 2. Set $\beta=1$.\\
\ \ \ 3. while \ $L_\sigma \{ \beta \zeta(\tilde{\zx}^{(i)}) + (1-\beta)
\zx^{(i)} \} >L_\sigma(\tilde{\zx}^{(i)}  )$ \\
\   \   \  \ \ \  \ \  \  $\beta=\gamma \beta$.\\
\   \   \  \ \ \  end\\
\ \ \ 4. $\zx^{(i+1)}=\beta \zeta(\tilde{\zx}^{(i)}) + (1-\beta) \tilde{\zx}^{(i)}$. Set $i=i+1$.\\
\ \ \ 5. If $ ||\tilde{\zx}^{(i)}-\tilde{\zx}^{(i-1)}||_2< \eta \sigma$ then $\sigma=\rho \sigma$.\\
{\bf while} \  $\sigma \ge \sigma_0$.\\
\hline
\end{tabular}
\end{table}
Here $\tilde{\zx}^{(i)}$ denotes the value of $\tilde{\zx}$ updated at the 
$i$ th iteration. The procedure of choosing  $\tilde{\zx}^{(0)}$ and $\sigma_{st}$ will be described in the next section. The value of $\lambda$ in \eqref{eq:opt2} controls the distance between $\zy$ and $\bA\zx_*(\sigma)$. A small value of $\lambda$ allows $\|\zy-\bA\zx_*(\sigma)\|_2^2$ to be larger. Note that according to \eqref{MEASUREMENT_EQUATION}, the term $\|\zy-\bA\zx\|_2^2$ is equal to the power of measurement noise. Hence, we should choose a small value for $\lambda$ when noise variance is larger. A procedure for choosing the value of $\lambda$ is described in \cite{l12}. The popular choice of $\lambda=0.1\|\bA^*\zy\|_{\infty}$. For minimizing
$L_{\sigma}$, we use $\zeta(\tilde{\zx})-\tilde{\zx}$ as the descent direction. The $\gamma$ is a standard backtracking line search parameter \cite{co}. The inner-iteration for minimizing $L_{\sigma}$
for a given $\sigma$ terminates when $||\tilde{\zx}^{(i+1)}-\tilde{\zx}^{(i)}||_2 < \eta \sigma$,
(see Step 5). We then update $\sigma=\rho\sigma$. The work in \cite{isl0} describes a procedure for choosing the values of $\eta$ and $\rho$. In particular, it has been shown that the ISL0 remains insensitive to the value of the parameters if we choose $\eta\in[0.1,0.7]$ and $\rho\in[0.2, 0.9]$. In this work, we set $\eta=0.5$ and $\rho=0.3$.
The stopping criterion of
ISL0 is based on a small value of $\sigma$ denoted by $\sigma_0$ which depends on the noise level. A wide range of numerical
simulations in noisy cases ($5$ dB to $20$ dB SNR) suggest that $\sigma_0=0.001$ is good choice. 

\subsection{The Handover algorithm}
The proposed handover algorithm starts with the $\ell_1$-optimization in Table-\ref{tab:algo2}. However, we allow the algorithm to provide only a rough estimation of $\zx$. Once a rough estimate $\hat{\zx}$ is obtained it is used as initial $\tilde{\zx}^{(0)}$ for ISL0 (see Table-\ref{tab:algo}). The value of $\sigma_{st}$ for ISL0 can be found in the following way. Assume that $\hat{\zx}$ is the minimizer of $L_{\sigma_{st}}(v)$ in \eqref{eq:opt2}. Then according to Lemma-\ref{lem1} and \eqref{8} we have
\begin{align}\label{eq:optst}
\hat{\zx}& = \zeta \{ \hat{\zx} \}\nonumber\\
&=\lambda 
\left[W_{\sigma}(\hat{\zx})/\sigma_{st}^2 +\lambda \bA^* \bA \right]^{-1} 
\bA^* \zy.
\end{align}
We need to solve \eqref{eq:optst} for $\sigma_{st}$. However, the equality in \eqref{eq:optst} may not hold in practice. Then the value of $\sigma_{st}^2$ can be approximated by:
\begin{align}\label{eq:optst2}
\sigma_{st}^2=\arg \min_{\sigma^2}\left\Vert\frac{W_{\sigma}(\hat{\zx})}{\sigma^2}\hat{\zx}-\lambda\bA^*(\zy-\bA\hat{\zx})\right\Vert_2^2
\end{align}
The optimization problem is non-convex, but one dimensional. An interior trust region algorithm \cite{trust} has been applied to estimate $\sigma_{st}$ from \eqref{eq:optst2} where the initial value of $\sigma$ is set to $\max_j(|[\hat{\zx}]_j|)$ (see \cite{jlzadoa} for justification).

Suppose the final output obtained from ISL0 is $\bar{\zx}$. Due to noise contribution in \eqref{MEASUREMENT_EQUATION}, the estimate $\bar{\zx}$ may not exact copy of $\zx$, and hence $\bar{\zx}$ will have spurious peaks. As a result, $\bar{\zx}$ needs thresholding to perform the code detection. We shall develop a thresholding procedure in the Section-\ref{sec:thre}. Let $\breve{\zx}$ be the thresholded vector constructed from $\bar{\zx}$. The BS can detect the active ranging codes and corresponding timing offsets from $\breve{\zx}$ by using the procedure described in Section-\ref{sec:srf}.
To obtain the estimate of channel impulse response (CIP), let us partition $\bar{\zx}=[ \ \bar{\zh}_1^{\intercal} \ \ \bar{\zh}_2^{\intercal} \ \ \cdots \ \ \bar{\zh}_G^{\intercal} \ ]^{\intercal}$. The estimate of CIP corresponding to the $\ell$-th active code is $\bar{\zh}_{\ell}$. 
 

\subsection{Computational Complexity Analysis}
In this section, we shall analyse computational complexity of the ISL0 algorithm. A similar procedure can be followed to analyze the complexity of the $\ell_1$ algorithm in Table-\ref{tab:algo2}.  As can be seen in Lemma-\ref{lem1}, the major fraction of the computation for ISL0 is involved 
in computing $\zeta(\tilde{\zx})$ in \eqref{8}. However
using the matrix inversion lemma in \cf{8} one
can verify that
\bea
&& 
\hspace{-1.1cm}
\zeta(\tilde{\zx}) = W_{\sigma}^{-1}(\tilde{\zx})\bA^*[ \bI/(\lambda \sigma^2) +  \bA W_{\sigma}^{-1}(\tilde{\zx}) \bA^*]^{-1} \zy.
 \label{10} 
\eea
We demonstrate a procedure to compute $\zeta(\tilde{\zx})$ in \eqref{10} efficiently by using FFT.  Let us rewrite \eqref{10} as
\begin{align}\label{eq:mod}
\zeta(\tilde{\zx})&=W_{\sigma}^{-1}(\tilde{\zx})\bA^*\hat{\zz}\\
\mathrm{where,} ~ \zy&=[\bR + \bI/ (\lambda \sigma^2)] \hat{\zz}, \nonumber\\
\bR&=\bA W_{\sigma}^{-1}(\tilde{\zx}) \bA^*\nonumber
\end{align}
Now partition $\tilde{\zx}$ into $G$ number of sub-vectors:
$$\tilde{\zx}=[\tilde{\zx}_1^{\intercal} \ \tilde{\zx}_2^{\intercal} \cdots \tilde{\zx}_G^{\intercal}]^{\intercal}$$
where length of each $\tilde{\zx}_i$ is $N_1$.
Construct the matrix $\hat{\bF}$ by extracting first $N_1$ columns of the Fourier matrix $\bF$ in \eqref{EXPRESSION_BS}.  We calculate \eqref{eq:mod} by using the following steps.
\begin{itemize}
\item At first we compute $\bR=\bA W_{\sigma}^{-1}(\tilde{\zx}) \bA^*$. Since $W_{\sigma}(\tilde{\zx})$ is a diagonal matrix, it follows using \cf{MEASUREMENT_EQUATION} and \eqref{ONE_TERMINAL_RELATION} that
\begin{align}
&\bA W_{\sigma}^{-1}(\tilde{\zx}) \bA^*=\nonumber\\
&\sum_{g=1}^G { [\Theta \ \mathrm{diag}( \Theta^{\intercal}  \zc_{g} )] \hat{\bF} W_{\sigma}^{-1}(\tilde{\zx}_g) \hat{\bF}^* [\Theta \ \mathrm{diag}( \Theta^{\intercal}  \zc_{g} )]^{\intercal}}\nonumber
\end{align}
 and 
\begin{align}\label{eq:step1}
[\hat{\bF}  W_{\sigma}^{-1}(\tilde{\zx}_g) \hat{\bF}^*]_{k, \ell}
&= \sum_{j=1}^{N_1} 
\zw_j(\tilde{\zx}_g) \erm^{- \irm 2 \pi (k-\ell) j/N }\nonumber\\
&= [\hat{\bF} \zw(\tilde{\zx}_g) ]_{k-\ell},
\end{align}
%
%
where $ \zw(\tilde{\zx}_g)$ is a vector constructed from the diagonal components of $W_{\sigma}^{-1}(\tilde{\zx}_g)$. 
We compute
$\hat{\bF} \zw(\tilde{\zx}_g)$  via FFT using $N \log_2(N)$ floating point operations. Recall that the entries of the code matrix $\bC$ are $\{+1,-1\}$. In addition $\Theta$ is a row selector matrix. Hence, a multiplication by  $[\Theta \ \mathrm{diag}( \Theta^{\intercal}  \zc_{g} )]$ does not require any floating point operation. In fact, constructing 
$$\bR_g := [\Theta \ \mathrm{diag}( \Theta^{\intercal}  \zc_{g} )] \hat{\bF} W_{\sigma}^{-1}(\tilde{\zx}_g) \hat{\bF}^* [\Theta \ \mathrm{diag}( \Theta^{\intercal}  \zc_{g} )]^{\intercal}$$ 
needs to extract a small block of $\hat{\bF} W_{\sigma}^{-1}(\tilde{\zx}_g) \hat{\bF}^*$, and changing the signs of some entries.
Finally, we compute $\bR=\sum_{i=1}^G \bR_{i}$ which requires $(G-1)(M+1)M/2$ flops. Hence this step requires $G \{ N \log_2(N) + M(M+1)/2 \} - M(M+1)/2$ flops in total.

\item 
Calculate $\hat{\zz}$ by solving $[\bR + 
\bI/ (\lambda \sigma^2)] \hat{\zz} = \zy$. By using Cholesky factorization, this  $M \times M$ 
positive definite system of equations need $O(\frac{1}{3}M^3)$ flops to compute $\hat{\zz}$.

%
\item Compute $\bA^*\hat{\zz}$ in parts, i.e. we partition
$\bA^*\hat{\zz}=[\tilde{\zz}_1 \ \tilde{\zz}_2 \ \cdots \tilde{\zz}_G]$, and form
\begin{align}\label{eq:step3}
\tilde{\zz}_g =
 \hat{\bF}^*[\Theta \ \mathrm{diag}( \Theta^{\intercal}  \zc_{g} )]^{\intercal}\hat{\zz}; \ \ g=1,2,\cdots G
 \end{align}
  by computing the IFFT of $[\Theta \ \mathrm{diag}( \Theta^{\intercal}  \zc_{g} )]^{\intercal}\hat{\zz}$. Recall that
forming $[\Theta \ \mathrm{diag}( \Theta^{\intercal}  \zc_{g} )]^{\intercal}\hat{\zz}$ does not require any additional floating point operation. Hence this step requires $O(GN\log_2(N))$ flops.
\item Finally, as
$W_{\sigma}^{-1}(\tilde{\zx})$ is diagonal, we need $GN_1$ multiplications to compute 
$\zeta(\tilde{\zx})$. 
\end{itemize}
Thus in total we need $O(2GN\log_2(N)+GM(M+1)/2+GN_1+1/3M^2(M-1.5))$ flops to compute 
$\zeta(\tilde{\zx})$. 
\subsection{Thresholding the recovered signal from ISL0}\label{sec:thre}
Let the final output obtained from ISL0 is $\bar{\zx}$. We denote
$\zv=\bar{\zx}-\zx.$ The vector $\zv$ can be viewed as the recovery error resulted due to noise. To perform a thresholding of $\bar{\zx}$, we analyze the statistical property of $\zv$. 
Using \eqref{8}, we can write
\begin{align}\label{eq:hy1}
\bar{\zx}&=\left[\frac{W_{\sigma}(\bar{\zx})}{\lambda\sigma^2} + \bA^* \bA \right]^{-1} 
\bA^* \zy.
\end{align}
Assuming $\|\zv\|_2$ is small, the first order Taylor series expansion of $W_{\sigma}(\bar{\zx})$ around $\zx$ is
\begin{align}
W_{\sigma}(\bar{\zx})=W_{\sigma}(\zx)-W_{\sigma}(\zx)\mathrm{diag}(\frac{\zx}{\sigma^2})\mathrm{diag}(\zv).
\end{align}
Now consider \eqref{eq:hy1},
\begin{align}
&\bA^* \zy=\left[\frac{W_{\sigma}(\bar{\zx})}{\lambda\sigma^2} + \bA^* \bA \right]\bar{\zx}\nonumber\\
&=\left[\frac{W_{\sigma}(\zx)}{\lambda\sigma^2}\left(\bI-\mathrm{diag}(\frac{\zx\zv}{\sigma^2})-\mathrm{diag}(\frac{\zx^2}{\sigma^2})\right) + \bA^* \bA \right]\zv\nonumber\\
&+\left[\frac{1}{\lambda\sigma^2}W_{\sigma}(\zx) +\bA^* \bA \right]\zx.
\end{align}
By ignoring second order terms in $\zv$ we have
\begin{align}
&\left[\frac{W_{\sigma}(\zx)}{\lambda\sigma^2}\left(\bI-\mathrm{diag}(\frac{\zx^2}{\sigma^2})\right) + \bA^* \bA \right]\zv\nonumber\\
&=
\bA^* \zy-\left[\frac{1}{\lambda\sigma^2}W_{\sigma}(\zx) +\bA^* \bA \right]\zx
\end{align}
For a small value of $\sigma$, we can neglect $W_{\sigma}(\zx)\frac{\zx}{\sigma^2}$, hence  
\begin{align}
\zv&=\left[\frac{W_{\sigma}(\zx)}{\lambda\sigma^2}\left(\bI-\mathrm{diag}(\frac{\zx^2}{\sigma^2})\right) + \bA^* \bA \right]^{-1}\bA^*\ze\nonumber\\
&=\bD\ze
\end{align}
where we define $\bD=\left[\frac{W_{\sigma}(\zx)}{\lambda\sigma^2}\left(\bI-\mathrm{diag}(\frac{\zx^2}{\sigma^2})\right) + \bA^* \bA \right]^{-1}\bA^*$. To compute $\bD$, we need the value of $\zx$ which is unknown in priori. Nevertheless, we can use an estimate of $\zx$ to compute $\bD$. In this work, we use $\bar{\zx}$ as an estimate of $\zx$. Furthermore, computing $\bD$ requires inverting a large size matrix. The computation task can be reduced significantly by applying matrix inversion lemma. Let us define $\bP=\frac{W_{\sigma}(\zx)}{\lambda\sigma^2}\left(\bI-\mathrm{diag}(\frac{\zx^2}{\sigma^2})\right)$. It can be verified that
\begin{align}
\bD=\bP^{-1}\bA^*\left[\bI+\bA\bP^{-1}\bA^*\right]^{-1}.
\end{align}
Let us partition the matrix $\bD$ such that $\bD^{\intercal}=[\bD^{(1)} \ \bD^{(1)} \cdots \bD^{(G)}]$ where each $\bD^{(i)}\in \mathbb{C}^{M\times N_1}$. Also partition $\zv=[\zv_1^{\intercal} \ \zv_2^{\intercal} \cdots \zv_G^{\intercal}]^{\intercal}$. Assume that $\ze$ is complex Gaussian with zero mean and a covariance matrix $\sigma^2\bI$. Hence, the entries of $\bD$ are independent of $\ze$. Then the variable $\|\zv_i\|_2^2=\|[\bD^{(i)}]^{\intercal}\ze\|_2^2$ has a generalized chi-square distribution of order $M$ (assuming $M<N_1$)\cite{chi2}. The procedure for computing the cumulative distribution function (CDF) of  a variable having generalized chi-square distribution has been described in \cite{chi1,chi2}. In this work, the CDF of $\|\zv_i\|_2^2$ for a threshold $\tau_i$ will be denoted by $\chi(\tau_i,\Lambda^{(i)},\sigma_e^2)$, where $\Lambda^{(i)}$ is the vector containing the singular values of $\left(\bD^{(i)} [\bD^{(i)}]^*\right)$.

Partition ${\zx}=[{\zx}_1^{\intercal} \ {\zx}_2^{\intercal} \ \cdots {\zx}_G^{\intercal}]^{\intercal}$ such that every ${\zx}_i\in \mathbb{C}^{N_1}$. Define a set 
$
S = \{ i : \| \zx_i\|_2 \ne 0 \}.$
Then for a given $i$ consider the two hypotheses:
$
[
\mathcal{H}_0 : i \notin S; \mathcal{H}_1 : i \in S
].$  
Note that under $\mathcal{H}_0$, the distribution of $\|\bar{\zx}_i\|_2^2$ is similar to the distribution of $\|\zv_i\|_2^2$. To perform hypothesis test on $\|\bar{\zx}_i\|_2^2$, we need to select a threshold parameter $\tau_i$. 
The procedure for selecting the value of $\tau_i$ will be described next. 
The value of $\|\bar{\zx}_i\|_2^2$ is checked against $\tau_i$ to take a decision between the two hypothesis:
\begin{align}
\|\bar{\zx}_i\|_2^2\gtrless^{\mathcal{H}_1}_{\mathcal{H}_0}\tau_i
\end{align}
The threshold $\tau_i$ is fixed to achieve a desired false alarm probability $\psi$ according to
\begin{align}\label{eq:thre}
\psi&=P(\|\bar{\zx}_i\|_2^2>\tau_i|\mathcal{H}_0)\nonumber\\
&=1-\chi(\tau_i,\Lambda^{(i)},\sigma_e^2)\end{align}
There are total $G$ number of sub-vectors i.e., $\{\bar{\zx}_i\}_{i=1}^G$. The overall false alarm probability can be defined:
\begin{align}\label{eq:fa}
P_{fa}=1-(1-\psi)^{G}
\end{align}
To perform thresholding of $\bar{\zx}$, we select a desired false alarm probability $P_{fa}$ first. For the $P_{fa}$, we can calculate the threshold parameter $\tau_i$ for every $\|\bar{\zx}_i\|_2^2$ by using \eqref{eq:thre}.

%

\section{Simulation Results}\label{sec:exp}
A typical $N=1024$ subcarrier OFDMA system, by following the WiMAX standards \cite{rang8,rang1},  has been chosen for the simulation.  In the system, the carrier frequency is $5.1$ GHz, and the associated sampling interval is $T_s=89.28$ ns. This corresponds to a subcarrier spacing of $10.94$ kHz. Length of the cyclic prefix is $64$ samples. Total $M=144$ subcarriers are reserved for the initial ranging purpose, and the total number of available random codes in matrix $\bC$ is $32$, i.e. $G=32$. 
The modulation pulse is a root-raised-cosine function with a roll-off $0.22$ and duration $10T_s$. The channel impulse response has a maximum order $P=30$, and the wireless cell radius is $2.5$ km, hence $D=186$. Similar to \cite{rang1,rang3}, we assume that BS has an approximate knowledge about $P_{\max}$ and we set $N_1=P_{\max}+D$ in \eqref{eq:el}.
The RTs follows a mixed channel model specified by ITU IMT-2000 standards: Ped-A, Ped-B, and Veh-A. The RTs select the channel models with equal probability. 
The mobile speed varies in the interval 
$[0,5]$ m/s for Ped-A, Ped-B channels, and $[5,20]$ m/s for Veh-A. Since the ranging signal is used to measure the system performance, the signal to noise ratio is defined as $\mathrm{SNR}=10\log_{10}\left(\frac{\sigma_h^2}{\sigma_e^2}\right)$, where $\sigma_h^2$ and $\sigma_e^2$ are the variances of channel impulse response $\zh$ and noise term $\ze$ respectively. Four different algorithms are considered for performance comparison. The proposed algorithm will be called ``Handover". The other three algorithms are the SMUD \cite{rang1}, SRMD scheme discussed in \cite{rang5}, and the MU-GLRT proposed in \cite{rang3}.

\begin{figure*}[]
\centerline{\subfloat[ ]{\includegraphics[width=8.5cm]{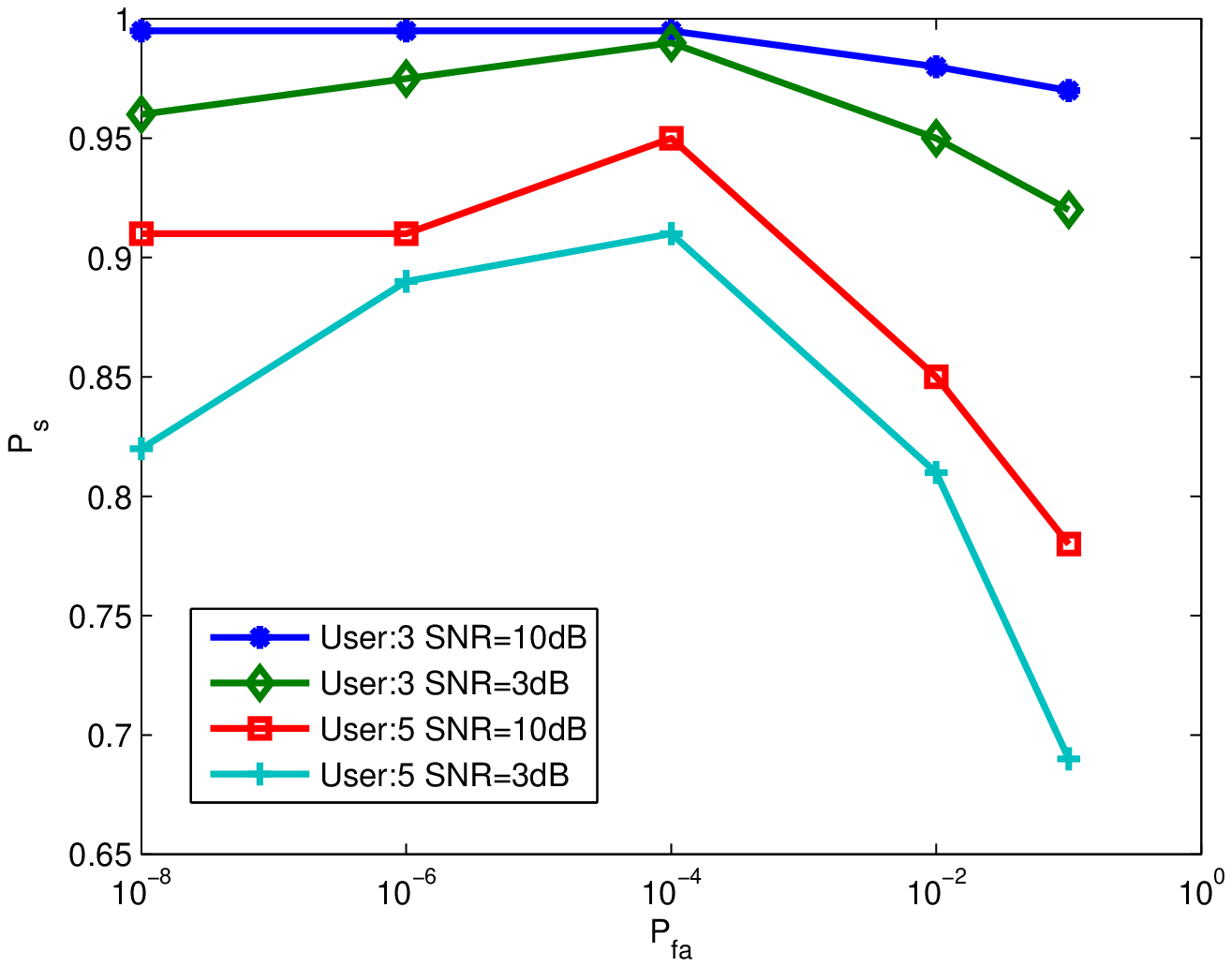}
\label{fig31}}
 \hfil \subfloat[ ]{\includegraphics[width=8.5cm]{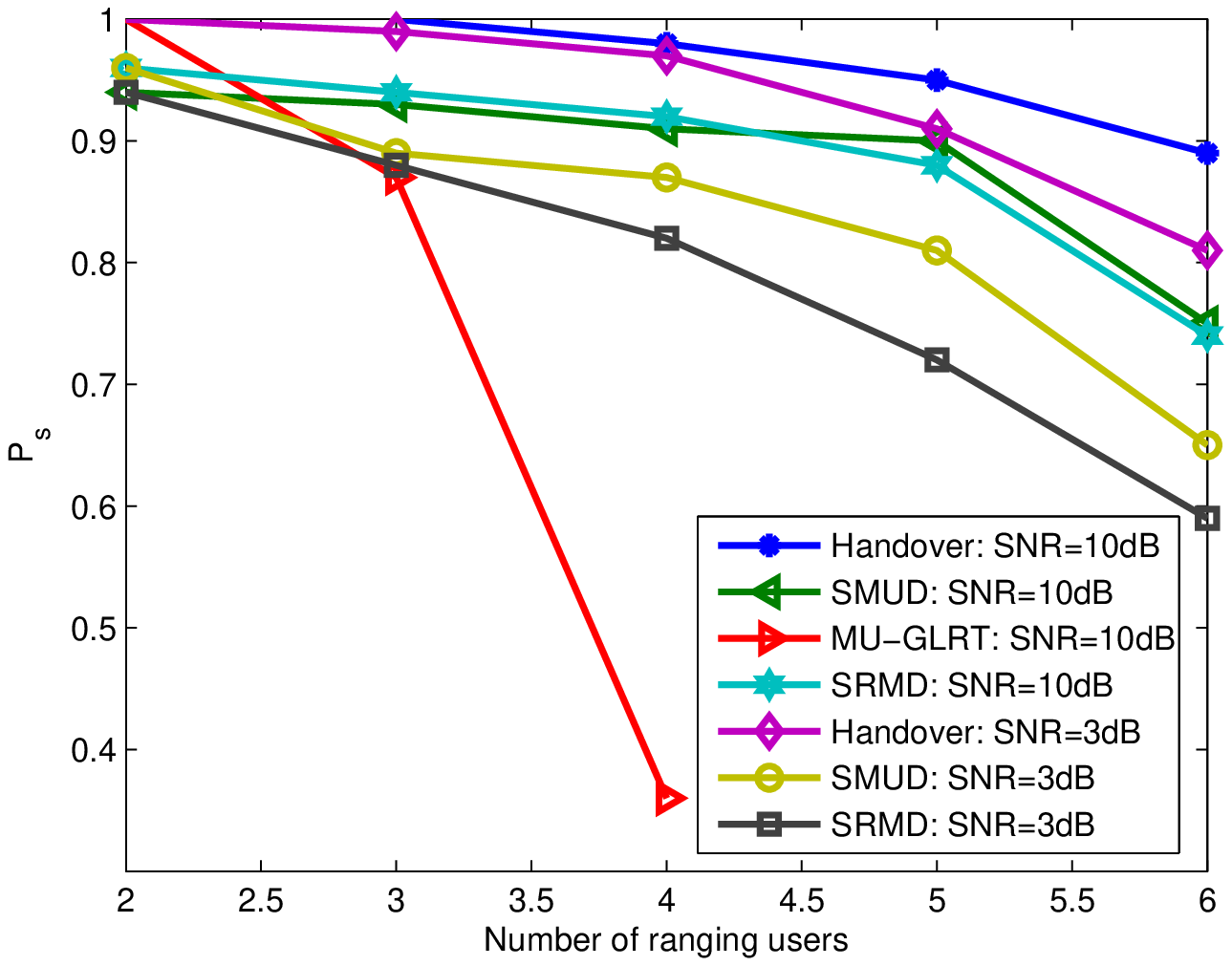}
\label{fig32}}} \caption{ Active code detection probability by differnt algorithms. (a) Code detection performance of the Handover algorithm for different values of false alarm probability $P_{fa}$. (b) Performance comparison of differnt algorithms, where the value of $P_{fa}=1e-4$ for Handover algorithm.} \label{fig3}
\end{figure*}

We start by finding a good choice of $\kappa$ to avoid 
unnecessary iterations in generating the rough estimate (to be
used as the initial guess by ISL0) via the $\ell_1$ optimization. From a wide range of simulations with different number of IR users and
SNR conditions we found that the performance of 
Handover remain almost same for $\kappa \ge 0.8$. Hence we recommend 
setting $\kappa=0.8$, which is used in all the following cases.
Figure \ref{fig3}(a) illustrates the code detection performance of the proposed algorithm for different values of false alarm probability $P_{fa}$ (see \eqref{eq:fa}). The performance is assessed in terms of success of code detection. 
Recall that the set of active IR code indices is $\mathcal{L}$.
Let $\hat{\mathcal{L}}$ be the set of code indices detected by an 
algorithm. The probability  that
${\mathcal{L}}=\hat{\mathcal{L}}$, denoted by $P_s$, is used 
to quantify the merit of the algorithm
We consider five different values of  $P_{fa}$ for code detection. As can be seen in Figure \ref{fig3}(a), the Handover algorithm provides optimum performance for $P_{fa}=1$e-4. Hence, we recommend setting $P_{fa}=1$e-4.

Figure \ref{fig3}(b) shows the code detection performance by different algorithms. The performance of MU-GLRT is worse for larger number of active ranging users compared to other three algorithms, whereas the Handover performs best. Note that at moderate SNR i.e., SNR$=10$ dB, the Handover algorithm can recover $6$ ranging users with high probability. The performance of SRMD is average compared to other algorithms. For instance, with $4$ users and SNR$=10$dB, the code detection probability of MU-GLRT, SRMD, SMUD and Handover are $0.36, 0.93, 0.91$ and $0.98$ respectively. The performance of MU-GLRT degrades rapidly with decreasing the SNR. Hence, we do not illustrate the result of MU-GLRT for lower SNR. With SNR$=3$dB and $4$ ranging users, the code detection probability of SRMD, SMUD and Handover are $0.88, 0.87$ and $0.97$ respectively. 
%

\begin{figure*}[]
\centerline{\subfloat[ ]{\includegraphics[width=8.5cm]{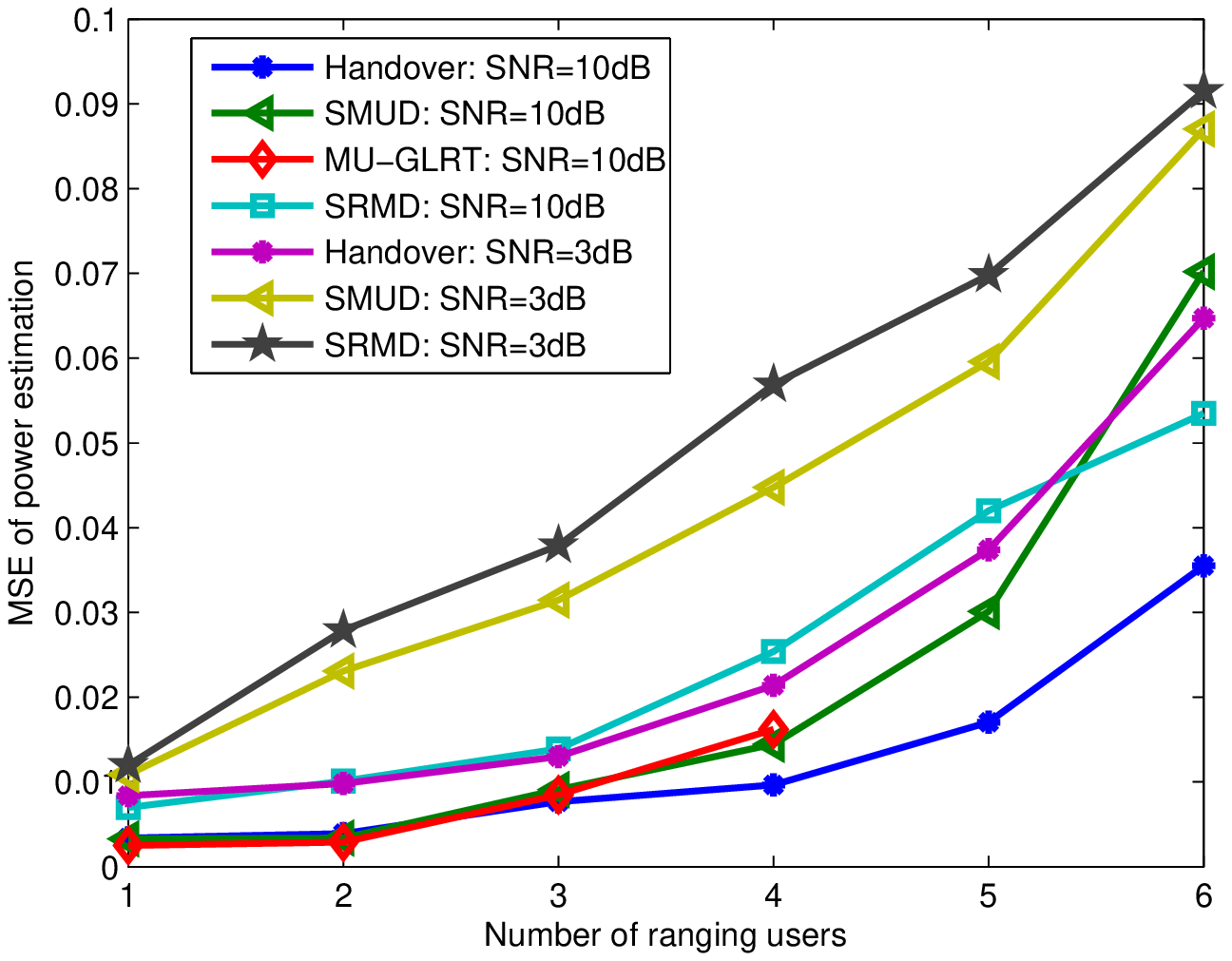}
\label{fig41}}
 \hfil \subfloat[ ]{\includegraphics[width=8.5cm]{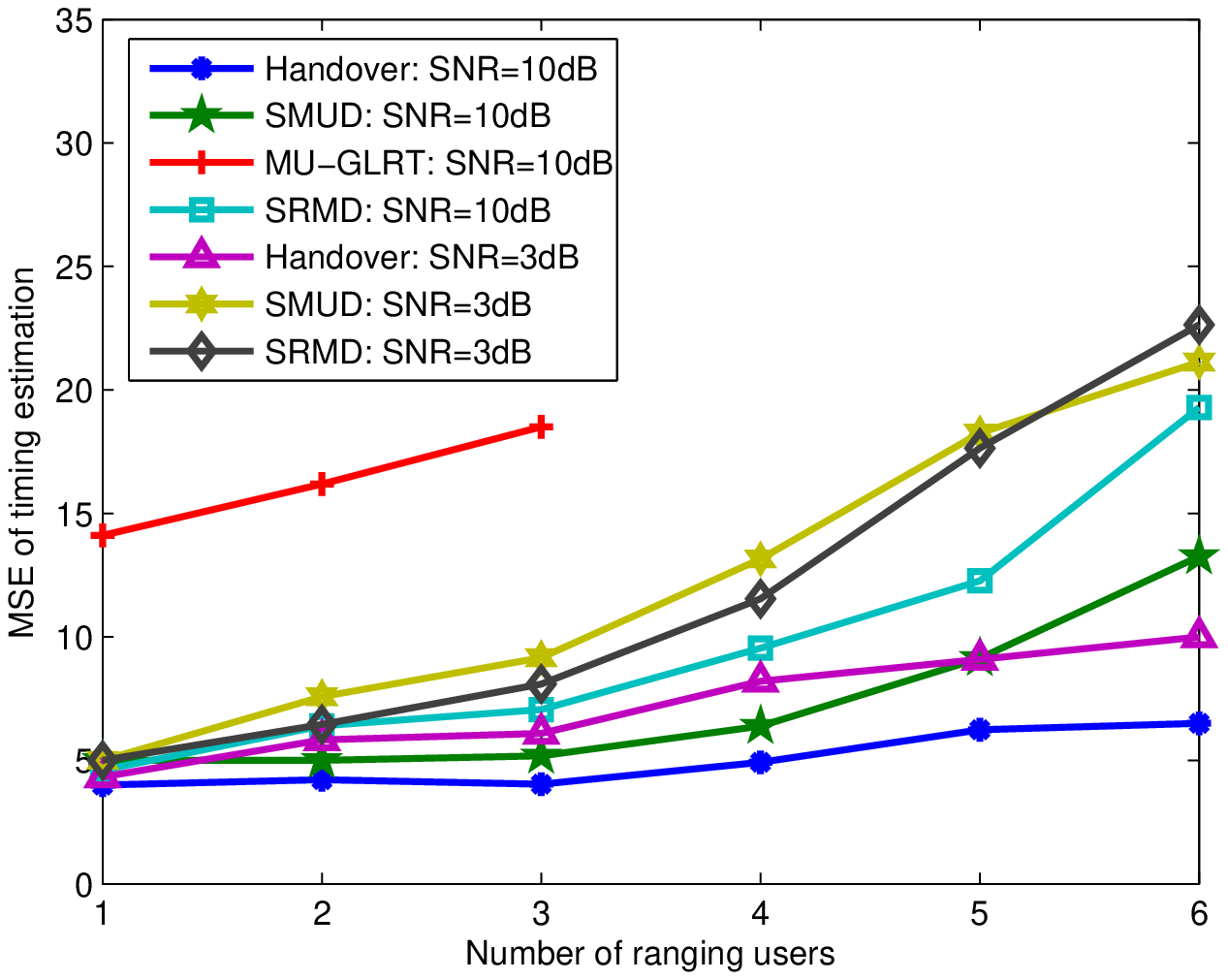}
\label{fig42}}} \caption{ MSE of ranging parameter estimations. (a) MSE of estimated channel power versus the number of active ranging users, (b) MSE of estimated timing offset versus the number of active ranging users. } \label{fig4}
\end{figure*}

Figure \ref{fig4} illustrates the accuracy of ranging parameter estimations by different algorithms. We do not compare the result with the MU-GLRT at low SNR as its performance is poor in the simulation environment. As can be seen in Figure \ref{fig4}(a), the MSE of power estimate increases with increasing the number of users. Note that with SNR$=10$dB, the SMUD, MU-GLRT and Handover exhibit similar performance for small number of users (i.e, for total users $2$ in Figure \ref{fig4}(a)). However, their performance difference increases with increasing the number of users. The MSE of power estimate from SMUD for $2$ and $5$ users (with SNR$=10$ dB) are $0.0033$ and $0.031$ respectively, whereas the MSE from Handover are $0.0034$ and $0.017$ respectively. 
The MSE of the timing estimates for the considered ranging algorithms are shown in Figure \ref{fig4}(b). As can be seen, the Handover algorithm outperforms other algorithms with big margin. For example, with $4$ ranging users and SNR$=10$dB, the MSE for Handover is $4.9$ which is $6.38$ and $9.54$ for SMUD and SRMD respectively. The MSE increases with increasing the number of users. For example, with $5$ ranging users and SNR$=10$ dB, the MSE of timing offset estimation by Handover, SMUD and SRMD are $6.24, 9.132$ and $12.27$ respectively. We see that the SMUD performs better than SRMD at high SNR ($10$ dB), however SRMD outperforms SMUD at low SNR (i.e, $3$ dB).  
We also compare computational complexity of the proposed algorithm with SMUD. The complexity of the SRMD algorithm has not been analysed in \cite{rang5}, hence we cannot incorporate the result in the figure.  With SNR= $10$ dB and total active users $6$, the Handover and SMUD requires $2.58$e7 and $1.64$e7 flops in average respectively to resolve the IR request.

\section{Conclusion}
In this work, we explored a formulation of the OFDMA initial ranging parameter estimation problem in a sparse signal representation framework. We started with developing a mathematical model that poses the ranging problem into a sparse signal recovery problem. An efficient procedure has been proposed that blends two different types of sparse recovery algorithms. The resulting algorithm exhibits efficient ranging parameter estimation performance. 

\bibliographystyle{IEEEtran}
\bibliography{rangj,Parj,Tcsb}

\end{document}